\documentclass[final,3p,times]{elsarticle}

\usepackage{amssymb}
\usepackage{amsmath}
\usepackage{mathtools}
\usepackage{lineno}
\usepackage[colorlinks=true, linkcolor=blue, citecolor=blue, urlcolor=blue]{hyperref}

\journal{Journal of Subatomic Particles and Cosmology}

\begin{document}
\begin{frontmatter}



\title{Production of Light Nuclei and Hypernuclei in Heavy-Ion Collisions}

\author[ucas]{Guannan Xie\texorpdfstring{\corref{cor1}}{}}
\ead{xieguannan@ucas.ac.cn}
\author[heidelberg]{Yue-Hang Leung\texorpdfstring{\corref{cor1}}{}}
\ead{leung@physi.uni-heidelberg.de}
\cortext[cor1]{Corresponding author}

\affiliation[ucas]{
  organization={University of Chinese Academy of Sciences},
  city={Beijing},
  postcode={101408},
  country={China}}

\affiliation[heidelberg]{
  organization={University of Heidelberg},
  city={Heidelberg},
  postcode={69120},
  country={Germany}}

\begin{abstract}
We review recent STAR and ALICE measurements of light-nucleus and hypernucleus yields, femtoscopic correlations, and collective flow presented at SQM 2026.
Statistical-hadronization calculations provide a useful baseline for integrated yields but do not simultaneously describe all measured light-nucleus ratios across collision energies and system sizes.
For bound states with mass number $A<4$, current coalescence calculations provide a broadly consistent description of yields, femtoscopic correlations, and collective flow, although the quantitative hypertriton comparison depends on the assumed few-body wave function. The suppressed production of resonant $^{4}$Li relative to compact $^{4}$He indicates an effect of nuclear structure and late-stage dynamics. However, the quantitative model comparison also depends on the treatment of feed-down from unstable states.
In high-multiplicity $p$+$p$ collisions, pion-deuteron femtoscopy further indicates that most observed (anti)deuterons are formed through nucleon fusion after strong decays of short-lived resonances. 
Taken together, these measurements show that production chronology and internal nuclear structure leave measurable imprints on the physics observables.
\end{abstract}



\begin{keyword}


 nuclei \sep hypernuclei \sep relativistic heavy-ion collisions \sep
coalescence \sep femtoscopy

\end{keyword}

\end{frontmatter}




\section{Introduction}
\label{sec:introduction}

High-energy nuclear collisions produce light (anti)nuclei and (anti)hypernuclei over a broad range of collision energies. Their production is intriguing because the binding energies of the lightest nuclei and hypernuclei are only a few MeV, far below the characteristic chemical-freeze-out temperature, which is of order
$100$ MeV over the collision energies considered. 
The simultaneous appearance of these weakly bound nuclei and an apparently thermal hadron population is often referred to as the ``snowballs in hell'' puzzle~\cite{BraunMunzinger:2015snowballs}. 
Meanwhile, this apparent tension opens an opportunity to investigate the formation dynamics of composite objects.

Two complementary physics goals motivate the measurements.  First, compound yield ratios such as $N_tN_p/N_d^2$ have been proposed as observables sensitive to baryon-density fluctuations and potentially to the QCD phase structure~\cite{Sun2018QCD,STAR2023Triton}. Second, hypernuclear yields and collective flow may constrain the strange sector of nuclear interactions, which is relevant to the hyperon puzzle in neutron stars~\cite{Lonardoni2015Hyperon}. In either case, a quantitative interpretation requires understanding when light nuclei and hypernuclei are formed, how they evolve and survive during the hadronic stage, and how their internal wave functions determine their production probability.

Measurements of light nuclei and hypernuclei have become an active research program at heavy-ion experiments worldwide. 
ALICE measures light nuclei, hypernuclei, and their antimatter counterparts at LHC energies, where the baryon chemical potential is close to zero, enabling precision studies of antimatter and nuclei formation at low baryon density. STAR's beam-energy-scan and fixed-target programs extend these measurements to the high-baryon-density region, reaching $\sqrt{s_{\mathrm{NN}}}=3$ GeV and $\mu_B \approx 720$ MeV, where the medium properties change drastically and the production of light nuclei and hypernuclei is significantly enhanced. The following discussion focuses on results presented at SQM 2026 and on two organizing questions: whether a common production picture can describe different observables, and whether the
yield retains measurable sensitivity to nuclear structure.

\section{Production frameworks}
\label{sec:models}

The statistical hadronization model (SHM) treats nuclei as additional hadronic species in chemical equilibrium.  Once the temperature, chemical potentials, and volume are fixed by ordinary hadrons, the primary yield of each nucleus is determined by its mass, spin degeneracy, conserved charges, and the adopted decay list~\cite{Andronic2011SHM}.  
The spatial size of the state does not explicitly enter. Consequently, the SHM provides a useful baseline when two states have similar masses but very different radii. Possible effects from decays of unstable nuclei and from hadronic evolution, including dissociation and regeneration, should be taken into account when interpreting comparisons with SHM predictions~\cite{Vovchenko2020Feeddown}.

In coalescence models, nuclei are formed by the fusion of constituents that are close in phase space near kinetic freeze-out. In the Wigner-function formalism, the production yield is proportional to the overlap integral of the constituent phase-space distributions and the nuclear Wigner function, and can be schematically written as 
\begin{equation}
 N_A \propto g_A
 \int \prod_{i=1}^{A} d^3r_i\,d^3p_i\,
 f_i(\boldsymbol{r}_i,\boldsymbol{p}_i)\,
 W_A(\{\boldsymbol{r}_{ij},\boldsymbol{p}_{ij}\}),
\label{eq:wigner}
\end{equation}
where $g_A$ is the spin degeneracy factor, $f_i(\mathbf{r}_i,\mathbf{p}_i)$ is the phase-space distribution of constituent $i$, and $W_A$ is the Wigner function of the nucleus. The production yield therefore depends on the source size, space-momentum correlations, and the nuclear wave function~\cite{Scheibl1999Coalescence}. A more spatially extended wave function generally has a smaller overlap with a compact emission source, although the production probability depends on the detailed structure of the nuclear wave function rather than its overall spatial extent alone.

Beyond these two approaches, quantum-molecular-dynamics approaches~\cite{Aichelin2020PHQMD} describe the dynamical formation of light nuclei through attractive interactions and identify bound configurations during the expansion, while kinetic approaches~\cite{Coci2023Kinetic} explicitly propagate formation and breakup reactions such as $\pi NN\leftrightarrow\pi d$ and $NNN\leftrightarrow Nd$. Unlike SHM and conventional coalescence models, these approaches explicitly describe the formation and subsequent evolution of light nuclei, allowing studies of their breakup and survival during the hadronic stage. Integrated yields alone may not uniquely distinguish different production scenarios, making differential observables and femtoscopic correlations valuable complementary probes.

\section{Yields and internal structure}
\label{sec:yields}

\subsection{Stable nuclei from RHIC to the LHC}

\begin{figure}[t]
\centering
\begin{minipage}[b]{0.35\linewidth}
  \centering
  \includegraphics[height=5.2cm]{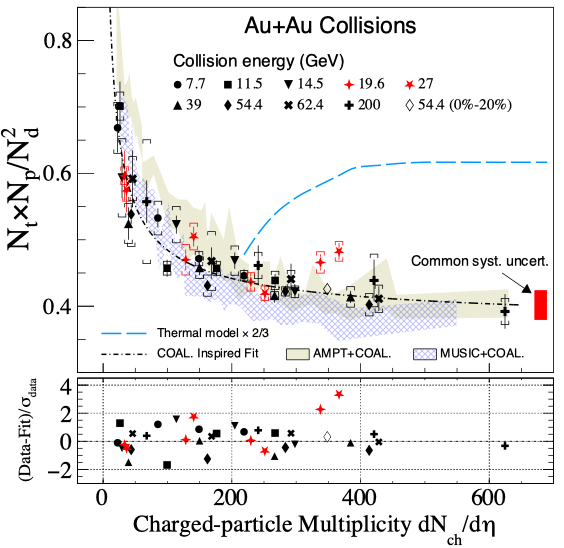}
\end{minipage}\hfill
\begin{minipage}[b]{0.61\linewidth}
  \centering
  \includegraphics[width=\linewidth]{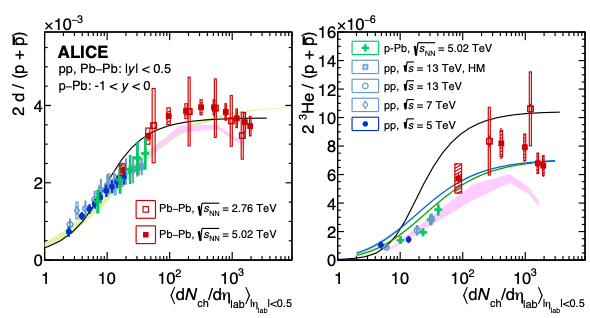}
\end{minipage}
\caption{Left: the compound ratio $N_tN_p/N_d^2$ versus charged-particle multiplicity in Au+Au collisions, compared with SHM and coalescence-based calculations~\cite{STAR2023Triton}. 
Right: $d/p$ and $^3$He$/p$ ratios at LHC energies, adapted from Fig. 7 of Ref.~\cite{ALICE:2022veq}.
The black curves show Thermal-FIST canonical-statistical-model calculations~\cite{Vovchenko:2018fiy}. The green curves denote two-body coalescence calculations; in the $^{3}He/p$ panel, the blue curve denotes three-body coalescence~\cite{sun:2018mqq}. The purple curves show UrQMD-hybrid calculations followed by coalescence for Pb–Pb collisions~\cite{Reichert:2022mek}.
}
\label{fig:stable}
\end{figure}

Before discussing the new SQM 2026 results, it is useful to briefly review earlier measurements of stable light nuclei, which provide important context for the recent developments. At RHIC Beam Energy Scan energies, the $d/p$ yield ratio is reasonably described by SHM calculations, whereas the measured $^3$He$/p$ and $t/p$ ratios are over-predicted by approximately a factor of two~\cite{STAR2024Nuclei}.  
For $^4$He, the comparison depends on the treatment of feed-down contributions from unstable nuclei, an effect that becomes increasingly important in the high-baryon-density region. A stringent test is provided by the compound ratio $N_tN_p/N_d^2$. As shown in Fig.~\ref{fig:stable} (left), the measured ratio decreases with charged-particle multiplicity over a broad range of collision energies. The coalescence calculations shown in the figure reproduce this trend within the current model and experimental uncertainties, whereas the corresponding SHM prediction increases with multiplicity and is inconsistent with the data~\cite{STAR2023Triton}.

At the LHC, a similarly non-uniform pattern emerges. The SHM describes the $A=2$ and $A=4$ yield ratios within uncertainties but overpredicts $A=3$, while the analytical coalescence calculations shown in Fig.~\ref{fig:stable} (right) provide a better description of the $A=2$ and $A=3$ ratios but underpredict $^4$He~\cite{ALICE2024Alpha}. 
The important conclusion is not that one production framework succeeds for every species. Rather, the combined RHIC and LHC measurements suggest that a more complete framework of light-nucleus production, beyond the current implementations of both SHM and coalescence, is required to consistently reproduce the observed yields across different nuclear species.

\subsection{Hypernuclear excitation functions and the strangeness population factor $S_{3}$}

\begin{figure}[t]
\centering
\begin{minipage}[b]{0.41\linewidth}
  \centering
  \includegraphics[width=\linewidth]{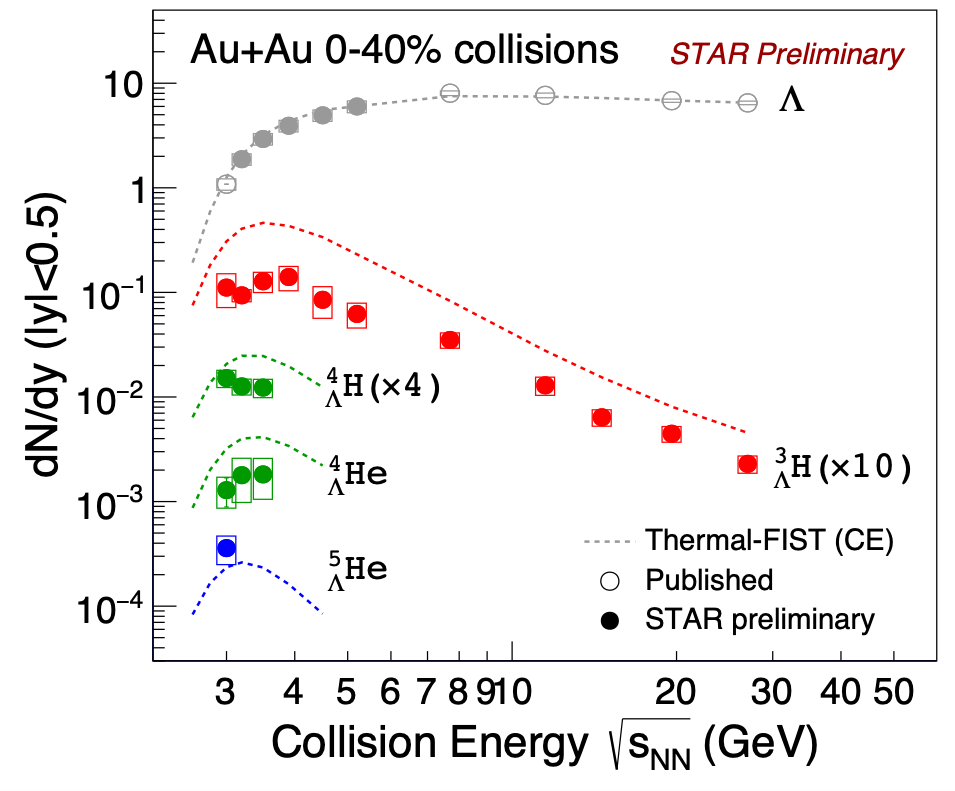}
\end{minipage}
\hspace{-0.015\linewidth}
\begin{minipage}[b]{0.40\linewidth}
  \centering
  \includegraphics[height=5.6cm]{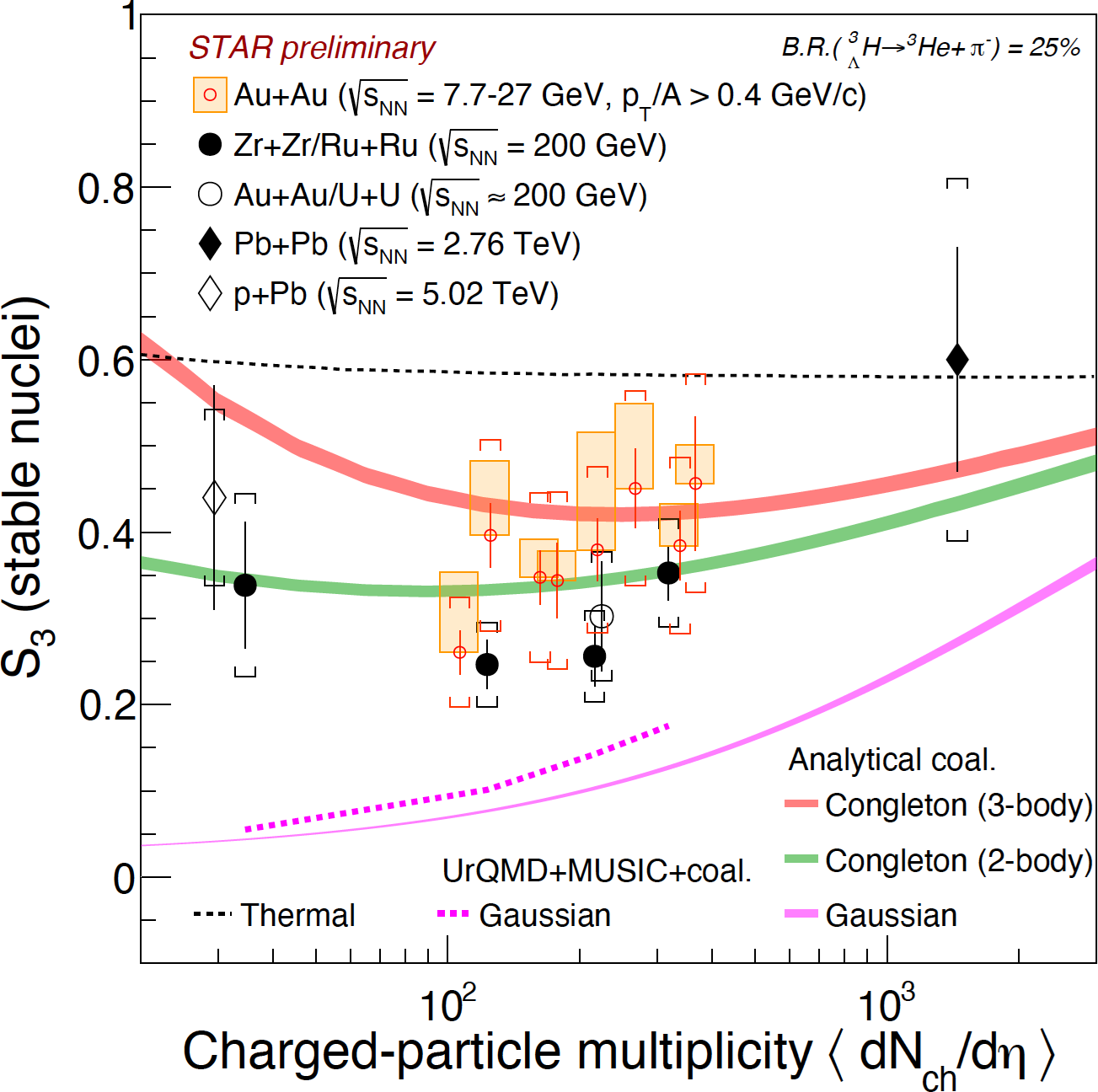}
\end{minipage}\hfill
\caption{Left: $\Lambda$ and hypernuclear excitation functions in Au+Au collisions compared with SHM calculations~\cite{STAR2022HyperYields,Zhou2025Hypernuclei}.  Right: $S_3$ versus charged-particle multiplicity compared with SHM and coalescence calculations~\cite{Leung2026DataGuided,STAR:2026ijb}.}
\label{fig:hyper}
\end{figure}

The hypernuclear excitation functions measured by STAR, shown in Fig.~\ref{fig:hyper} (left), illustrate the interplay between strange-particle production and hypernucleus formation. The $^{3}_{\Lambda}$H yield reaches a maximum around $\sqrt{s_{\mathrm{NN}}}=3$--4 GeV and decreases at higher energies. Toward lower collision energies, the increasing baryon density enhances the probability of hypernucleus formation, while canonical strangeness suppression acts in the opposite direction. The canonical-ensemble SHM calculation based on Thermal-FIST, shown in Fig.~\ref{fig:hyper} (left), overpredicts the measured hypertriton (${}^{3}_{\Lambda}$H) yields over the explored energy range, similar to its behavior for $^3$He and $t$. The discrepancy decreases for the more tightly bound $A=4$ hypernuclei, while, perhaps surprisingly, the measured $A=5$ hypernucleus yield is consistent with the calculation within uncertainties~\cite{STAR2022HyperYields,Zhou2025Hypernuclei}. This ordering may be related to the internal structure of the hypernuclei, which can influence both their formation and their survival during the hadronic stage. Disentangling these effects requires dynamical models that couple the formation and dissociation of (hyper)nuclei.

The strangeness population factor,
\begin{equation}
 S_3 =
 \frac{{}^{3}_{\Lambda}\mathrm{H}/{}^{3}\mathrm{He}}
 {\Lambda/p}
\label{eq:s3}
\end{equation}
removes the different proton and $\Lambda$ abundances and permits a more direct comparison of $^{3}_{\Lambda}$H and $^3$He formation. As shown in Fig.~\ref{fig:hyper} (right), measurements from different collision systems and energies mostly lie in the range $S_3\simeq0.3$--0.5~\cite{STAR:2026ijb}, well below unity, qualitatively consistent with coalescence expectations based on the much larger spatial extent of the $^{3}_{\Lambda}$H compared with $^3$He. Both the data-guided coalescence model~\cite{Leung2026DataGuided} and a hybrid calculation combining UrQMD+MUSIC with coalescence~\cite{2021UrqmdCoalescnece} underestimate the measured $S_3$ values when a simple Gaussian $^{3}_{\Lambda}$H wave function is used, whereas more realistic Congleton-type wave functions provide a significantly better description.
Although these wave functions reproduce similar rms radii, the Congleton-type wave functions have a substantially larger short-distance component, leading to a greater overlap with the emission source.
These comparisons suggest that $^{3}_{\Lambda}$H yields and yield ratios provide sensitivity to the short-distance structure of the $^{3}_{\Lambda}$H wave function, and therefore to the underlying $\Lambda N$ interaction.

\section{Unstable nuclei}
\label{sec:unstable}

\begin{figure}[t]
\centering
\begin{minipage}[b]{0.46\linewidth}
  \centering
  \includegraphics[height=5.6cm]{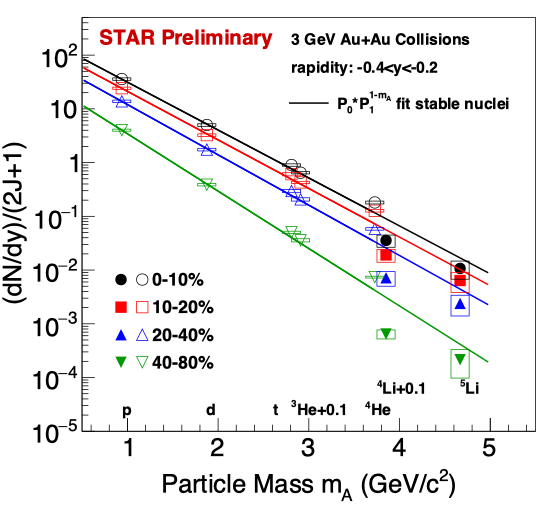}
\end{minipage}
\hspace{-0.015\linewidth}
\begin{minipage}[b]{0.46\linewidth}
  \centering
    \includegraphics[height=6.0cm]{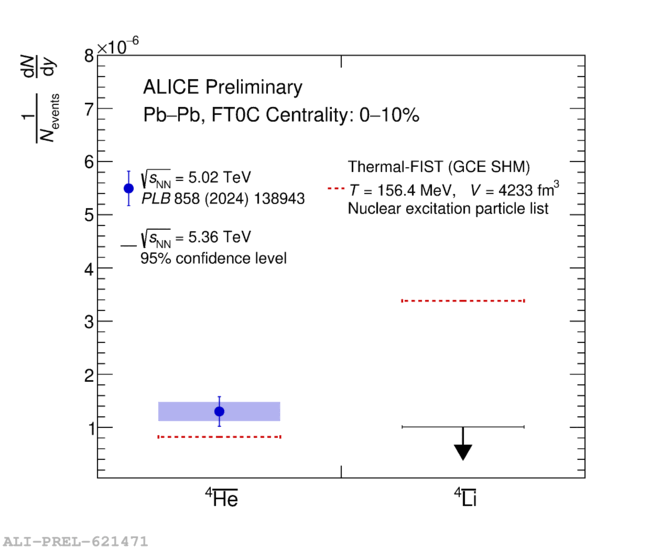}
\end{minipage}
\caption{Left: Spin-degeneracy-scaled light-nucleus yields as a function of mass in $\sqrt{s_{\rm{NN}}}=3$ GeV Au+Au collisions, including preliminary $^4$Li and $^5$Li results presented at SQM 2026~\cite{HuWu2026SQM}. The lines are exponential fits to the stable nuclei yields. Right: the upper limit on the $^4\overline{\mathrm{Li}}$ yield in $\sqrt{s_{\rm{NN}}}=5.36$ TeV Pb+Pb collisions, compared with the measured $^4\overline{\mathrm{He}}$ yield and the SHM expectation~\cite{Lucia2026SQM}.}
\label{fig:lithium}
\end{figure}

Near-threshold resonances provide an incisive comparison because their masses can be close to those of stable nuclei while their widths and spatial structures differ. 
STAR reconstructs $^4$Li and $^5$Li through the $p$--$^3$He and $p$--$^4$He correlation functions, respectively. 
The correlation analysis combines the relevant partial waves, low-energy phase-shift information, Coulomb effects, and the femtoscopic source to isolate the resonant contribution. 
In $\sqrt{s_{\mathrm{NN}}}=3$ GeV Au+Au collisions, the spin-degeneracy-scaled $^4$Li yield is found to be substantially lower than the $^4$He yield within the current experimental uncertainties, and an SHM calculation overpredicts the measured $^4$Li yield~\cite{HuWu2026SQM}, shown in Fig.~\ref{fig:lithium} (left). 
Without feed-down from unstable nuclei, the calculation slightly underpredicts the measured $^4$He yield, while including feed-down leads to an overprediction.
Thus, the suppression of the spin-degeneracy-scaled $^4$Li yield relative to $^4$He is a feature of the measurement, whereas the quantitative level of agreement with the SHM depends on the feed-down contribution assumed for the measured $^4$He yield.

ALICE has performed the antimatter analogue using $\bar{p}$--$^3\overline{\mathrm{He}}$ femtoscopy in 5.36 TeV Pb+Pb collisions, shown in Fig.~\ref{fig:lithium} (right). 
A candidate $^4\overline{\mathrm{Li}}$ signal is observed with a local significance of $2.1\sigma$, and the resulting upper limit lies well below the SHM expectation, while the compact $^4\overline{\mathrm{He}}$ yield is described~\cite{Lucia2026SQM}.  
The STAR $^4$Li and ALICE $^4\overline{\mathrm{Li}}$ measurements therefore point in the same direction: resonant $A=4$ lithium states are suppressed relative to their compact helium counterparts.  
The STAR measurements further suggest that this suppression may reflect not only the spatially extended nature of $^4$Li but also the relatively small partial-wave scattering cross section of the $p$--$^3$He system compared with $p$--$^4$He, which forms $^5$Li. 

\section{Femtoscopic correlations and collective flow}
\label{sec:dynamics}

\begin{figure}[t]
\centering
\begin{minipage}[b]{0.33\linewidth}
  \centering
  \includegraphics[width=\linewidth]{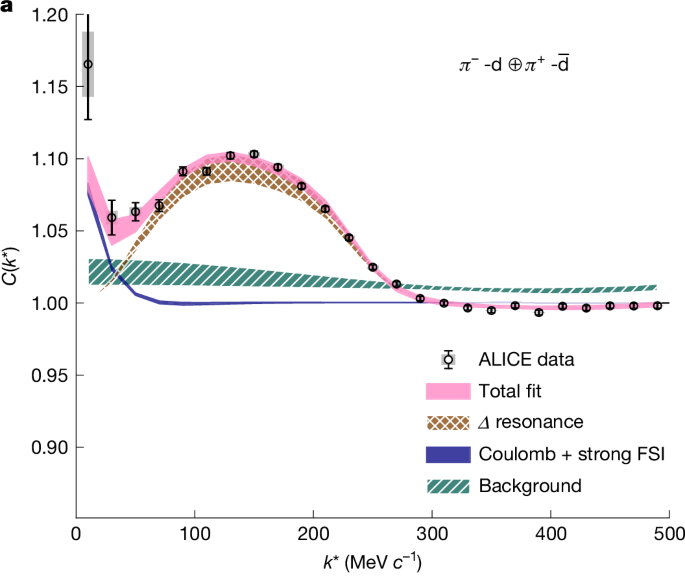}
\end{minipage}\hfill
\begin{minipage}[b]{0.31\linewidth}
  \centering
  \includegraphics[width=\linewidth]{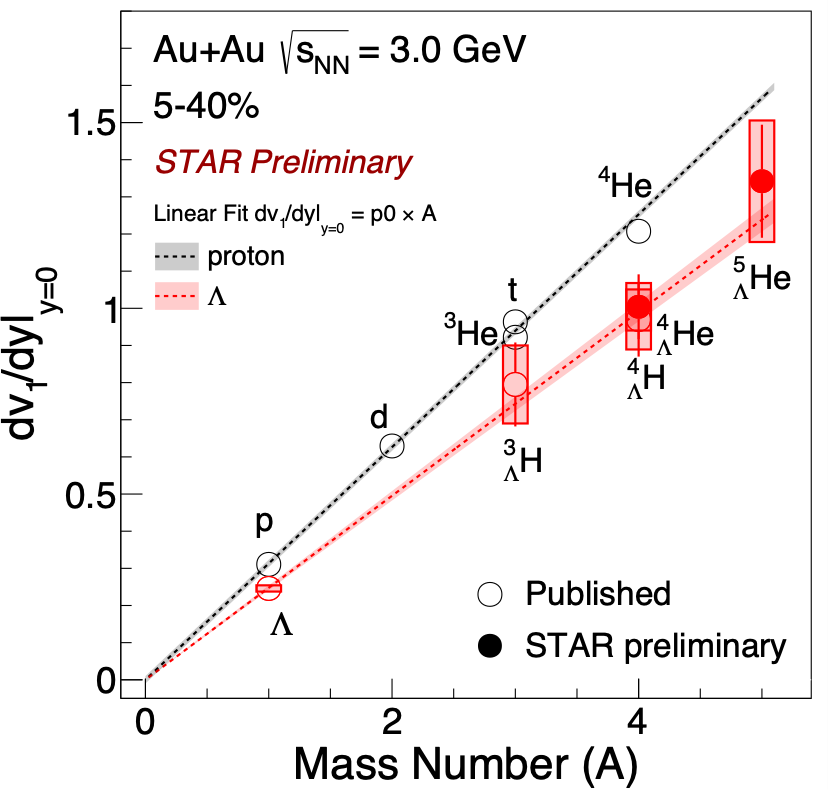}
\end{minipage}\hfill
\begin{minipage}[b]{0.33\linewidth}
  \centering
  \includegraphics[width=\linewidth,height=1.0\linewidth]{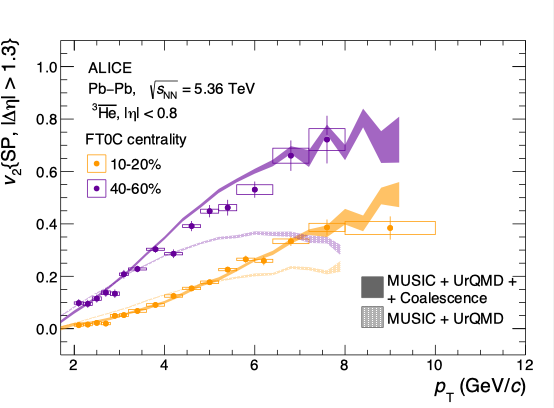}
\end{minipage}
\caption{Left: $\pi^+$--$d$ and $\pi^-$--$d$ correlation functions measured in $\sqrt{s}=13$ TeV $p$+$p$ collisions  together with the corresponding fit function. The brown cross-hatched band represents contributions from the $\Delta$ resonance, the blue band denotes the Coulomb and strong FSI interactions, and the teal diagonally hatched band corresponds to the residual background~\cite{ALICE2025Deuteron}.  
Middle: Nuclei and hypernuclei $v_{1}$ slopes at midrapidity in $\sqrt{s_{\rm{NN}}}=3$ GeV Au+Au collisions~\cite{STAR:2021ozh,STAR2023HyperFlow,Zhou2025Hypernuclei}. The lines represent linear fits through the origin. Right: $^3\overline{\mathrm{He}}$ $v_{2}$ in $\sqrt{s_{NN}}=5.36$ TeV Pb+Pb collisions, compared with calculations with and without coalescence~\cite{ALICE2026Flow}.}
\label{fig:dynamics}
\end{figure}

Femtoscopic correlations provide a more direct probe of the production chronology than integrated yields. ALICE measured $\pi^\pm$--$d$ correlations in high-multiplicity $\sqrt{s}=13$ TeV $p$+$p$ collisions and observed the residual correlation expected from $\Delta(1232)$ decays, shown in Fig.~\ref{fig:dynamics} (left). A deuteron produced independently of the pion does not preserve this parent-resonance correlation, whereas nucleon coalescence of a nucleon from the $\Delta$ decay with another nearby nucleon does. Using a data-driven approach together with the measured contributions from strong resonances, ALICE estimated that $(88.9\pm6.3)\%$ of the observed (anti)deuterons are formed through nucleon fusion involving resonance-decay nucleons~\cite{ALICE2025Deuteron}. This provides direct evidence that deuteron formation predominantly occurs after the decay of short-lived resonances, offering a natural explanation for how such weakly bound nuclei can emerge despite the much hotter medium present at earlier times.

Collective flow provides a complementary probe of the production mechanism. In $\sqrt{s_{\rm{NN}}}=3$ GeV Au+Au collisions, the midrapidity directed-flow slopes of nuclei and hypernuclei approximately follow mass-number scaling upto $A=5$~\cite{STAR:2021ozh,STAR2023HyperFlow,Zhou2025Hypernuclei}, shown in Fig.~\ref{fig:dynamics} (middle).  
Such scaling is naturally expected if light nuclei and hypernuclei inherit the collective motion of their constituent baryons during formation. At the LHC, the measured $^3\overline{\mathrm{He}}$ elliptic flow is described by hydrodynamic calculations coupled to a coalescence afterburner, while a calculation without the coalescence step underestimates the data at high-$p_{\mathrm{T}}$ as shown in Fig.~\ref{fig:dynamics} (right). The same dataset also provides the first observation of significant ${}^{3}_{\Lambda}\rm{H}$ elliptic flow in heavy-ion collisions. Within the current uncertainties, the measured ${}^{3}_{\Lambda}\mathrm{H}$ $v_2$ is consistent with that of $^3\overline{\mathrm{He}}$ and with hydrodynamic calculations coupled to a coalescence afterburner~\cite{ALICE2026Flow}.

\section{Summary and outlook}
\label{sec:summary}

The measurements reviewed here expose complementary strengths and limitations of statistical hadronization and coalescence. Statistical hadronization provides an economical baseline for many integrated yields, but pion–deuteron femtoscopy shows that most observed (anti)deuterons form after short-lived-resonance decays, disfavoring a literal interpretation of their direct formation at chemical freeze-out. Coalescence broadly describes the yield, flow, and femtoscopic systematics for (hyper)nuclei with $A<4$, although hypertriton predictions depend on the assumed few-body wave function. For $A\geq4$, the remaining tensions may require dynamical formation and breakup, but their significance depends particularly on feed-down from unstable states and does not establish a model-independent breakdown at a particular mass number.

Comparisons among states of similar mass but different structure make yields and ratios sensitive to nuclear wave functions. The measured $S_3\simeq0.3$--0.5 and the suppression of ${}^{4}$Li relative to ${}^{4}$He are qualitatively consistent with this picture, while quantitative conclusions remain model dependent. Single- and double-$\Lambda$ hypernuclei probe the $YN$ and $\Lambda\Lambda$ interactions relevant to neutron-star matter. 
At the EIC, spectator-tagged deep-inelastic scattering with deuteron and ${}^{3}$He beams can independently constrain few-body wave functions used in coalescence calculations~\cite{Cosyn:2020kwu,Friscic:2021oti}, while next-generation high-statistics heavy-ion programs at FAIR, HIAF, and NICA will enable precision measurements of these systems, providing new opportunities to connect hypernuclear spectroscopy with the production dynamics discussed in this review.

\section{Acknowledgments}
\label{sec:ack}

This work was supported in part by the National Natural Science Foundation of China (Grant No. 12305146) and by the CAS Project for Young Scientists in Basic Research (Grant No. YSBR-088).



\bibliographystyle{elsarticle-num}
\bibliography{sqm2026_template}



\end{document}